\documentstyle[preprint,aps,epsf,tighten]{revtex}	
\def\aprle{\buildrel < \over {_{\sim}}}

\begin{document}
%
%
\preprint{$
\begin{array}{l}
\mbox{BA-04-06}\\
\mbox{FERMILAB-Pub-04/043-T}\\
\mbox{April 2004}\\[0.5in]
\end{array}
$}
\title{Resonant leptogenesis in a predictive $SO(10)$ grand unified model\\}
\author{Carl H. Albright}
\address{Department of Physics, Northern Illinois University, DeKalb, IL
60115\\
       and\\
Fermi National Accelerator Laboratory, P.O. Box 500, Batavia, IL
60510\footnote{electronic address: albright@fnal.gov}}
\author{S.M. Barr}
\address{Bartol Research Institute,
University of Delaware, Newark, DE 19716\footnote{electronic address:
smbarr@bartol.udel.edu}}

\maketitle
\begin{abstract}

An $SO(10)$ grand unified model considered previously by the authors featuring 
lopsided down quark and charged lepton mass matrices is successfully predictive
and requires that the lightest two right-handed Majorana neutrinons be nearly 
degenerate in order to obtain the LMA solar neutrino solution.  Here we use 
this model to test its predictions for baryogenesis through resonant-enhanced 
leptogenesis.  With the conventional type I seesaw mechanism, the best 
predictions for baryogenesis appear to fall a factor of three short of the 
observed value.  However, with a proposed type III seesaw mechanism leading 
to three pairs of massive pseudo-Dirac neutrinos, resonant 
leptogenesis is decoupled from the neutrino mass and mixing issues with  
successful baryogenesis easily obtained.

\thispagestyle{empty}
\end{abstract}
\vskip .2in
PACS numbers: 14.60.Pq, 12.10.Dm, 12.15Ff, 12.60.Jv\\
%
%

\section{INTRODUCTION}

\indent The accumulation of refined data on both neutrino oscillation and 
quark flavor physics has posed increasingly severe tests for quark and lepton
models which attempt to explain the mass and mixing data.  This is especially
true for Grand Unified models which must relate both the quark and lepton 
sectors. In fact, at present several such models \cite{models} do survive 
these tests given the present levels of experimental precision for the 
Cabibbo-Kobayashi-Maskawa (CKM) quark and Maki-Nakagawa-Sakata (MNS) 
lepton mixing matrix elements.  Two other hurdles for these Grand Unified 
models involve the issues of flavor-changing neutral currents, for example in 
$\mu \rightarrow e + \gamma$ and $\tau \rightarrow \mu + \gamma$ decays, and 
the survival of baryogenesis in the early Universe.  It is the latter hurdle
which we wish to address in this paper.

It appeared early on that Grand Unified Theories (GUTs) provided the necessary
conditions for successful baryogenesis as postulated by Sakharov 
\cite{sakharov}: 
baryon-violating interactions which involve both $C$- and $CP$-violation and
which occur out-of-equilibrium in the early Universe.  Since then we have 
learned that the baryon-violating interactions which occur near the GUT scale
conserve $B - L$, so that any net baryon number generated by them can be 
washed out by sphaleron $B + L$ interactions which 
occur in thermal equilibrium with the expanding universe.  Only if the net
baryon number is generated with $\Delta (B - L) \neq 0$ interactions will it 
not be erased by the sphaleron interactions \cite{ht}.

For this reason, Fukugita and Yanagida suggested that leptogenesis may play
a necessary primary role for baryogenesis \cite{fy}.  An excess in lepton 
number generated by the lepton-violating decays of the heavy 
right-handed neutrinos can be converted into a baryon excess by sphaleron 
interactions in thermal equilibrium above the critical electroweak 
symmetry-breaking temperature. 
In this decay or direct ``$\,\epsilon'\,$'' $CP$-violating 
scenario, the $CP$ violation is generated through an interference between 
the decay tree graph and the absorptive part of the one-loop decay vertex.
This mechanism requires very heavy $N_i,\ i = 1,2,3$ right-handed neutrinos,
with the lightest being heavier than roughly $10^9$ GeV. Thus with a 
moderate hierarchy of heavy right-handed neutrinos and with the lightest 
satisfying the above bound, satisfactory leptogenesis can be generated.

However, it has been observed that successful leptogenesis may also arise 
through indirect ``\,$\epsilon$\,'' $CP$-violating mass-mixing effects in 
the decays of two quasi-degenerate heavy right-handed neutrinos \cite{res}.  
In other words, resonant enhancement of the lepton asymmetry can be generated 
if the two neutrinos are sufficiently close in mass that level crossing can 
occur. Ellis, Raidal, and Yanagida as well as Akhmedov, Frigerio and Smirnov 
have performed phenomenological studies \cite{eryafs} of neutrino mass 
matrices and shown that sufficient lepton asymmetry can be generated in the 
case that the level crossing involves the two lighter right-handed neutrinos, 
$N_1$ and $N_2$; moreover, the quasi-degenerate masses can lie as low as 
$10^8$ GeV, at least a factor of 10 below the bound obtained in the generic 
hierarchical case.  Recently Pilaftsis and Underwood have proposed a model
of the lepton sector in which the nearly degenerate neutrino pair is as 
light as 1 TeV \cite{pu}.

It is important to test this suggested scenario in a realistic model of 
quark and lepton masses in as quantitative a manner as possible.  At issue 
is whether or not the desired results for both leptogenesis and neutrino 
oscillations can be obtained simultaneously for the model in question.  
In this paper we explore leptogenesis with such resonant enhancement in 
an $SO(10)$ GUT model proposed several years ago by us in collaboration 
with Babu \cite{abb} with a series of refinements by us since then \cite{ab}.  
This model is numerically very predictive and is found to explain accurately 
the present data on quark masses and mixings as well as the large mixing angle 
(LMA) Mikheyev-Smirnov-Wolfenstein (MSW) \cite{msw} solar neutrino 
solution with near maximal atmospheric neutrino mixing.
The lopsided texture of the 2-3 submatrices of the down quark and charged 
lepton mass matrices neatly explains the small mixing in that sector of 
the CKM matrix, i.e., small $V_{cb}$ and $V_{ts}$, as well as the near maximal 
mixing in that sector of the MNS matrix \cite{mns}, i.e., large $U_{\mu 3}$.
The LMA solar neutrino solution arises from the type I seesaw mechanism as an
interplay between the Dirac neutrino matrix and the right-handed Majorana
mass matrix, $M_R$. In fact, the 2-3 submatrix of $M_R$ has a zero determinant
which nearly replicates the corresponding sector for the right-handed 
Dirac neutrino matrix product, $M_N^\dagger M_N$.  With the full structure of 
$M_R$ spelled out to give the LMA solution, one finds the lighter two 
masses $N_1$ and $N_2$ are nearly degenerate, so at least the possibility 
of resonant enhancement of leptogenesis exists in this model.   
We shall examine how successful the model can be in achieving leptogenesis 
sufficient to generate the desired amount of baryogenesis observed in Nature.  

As an alternative approach, we can reformulate the model with the type III
seesaw mechanism developed in \cite{typeIII,typeIIIab}. Whereas in the usual 
type I see-saw mechanism there are three superheavy Majorana neutrinos, 
in the type III see-saw, there are six superheavy neutrinos that typically
form three pseudo-Dirac pairs. The desired resonant enhancement of leptogenesis
can result from the mixing of the two neutrinos of the lightest pseudo-Dirac 
pair. In this scheme, the LMA solar neutrino mixing solution emerges rather 
naturally without fine-tuning of the Dirac and Majorana neutrino matrices, and 
the heavy lepton decay asymmetry is effectively decoupled from the light 
neutrino neutrino mass and mixing issues. As a result we shall see that 
satisfactory leptogenesis is easily obtained in this scenario.

In Sect. II we summarize briefly the features of the $SO(10)$ model and take 
the opportunity to update the input parameters so as to give even better 
agreement with the quark and lepton mass and mixing data.  The formulas 
relevant for leptogenesis involving two quasi-degenerate right-handed 
neutrinos are summarized in Sect. III.  We then apply these formulas to 
the model, modify the model slightly and compare the  leptogenesis 
and oscillation results with experimental information.  In Sect. IV
the necessary formulas are presented for the type III seesaw and numerically 
satisfactory examples are given.  Conclusions follow in Sect. V.

\section{$SO(10)$ Model with $U(1) \times Z_2 \times Z_2$ Flavor Symmetry}

The GUT model in question \cite{abb,ab} is based on the grand unified group 
$SO(10)$ with a $U(1) \times Z_2 \times Z_2$ flavor symmetry.  The model 
involves a minimum set of Higgs fields which solves the doublet-triplet 
splitting problem.  This requires just one ${\bf 45}_H$ whose VEV points in 
the $B-L$ direction, two pairs of ${\bf 16}_H,\ {\bf \overline{16}}_H$'s 
which stabilize the solution, along with several Higgs fields in the 
${\bf 10}_H$ representations and Higgs singlets \cite{br}.  The Higgs 
superpotential exhibits the $U(1) \times Z_2 \times Z_2$ symmetry which is 
used for the flavor symmetry of the GUT model.  The combination of VEVs, 
$\langle {\bf 45_H}\rangle_{B-L},\ \langle 1({\bf 16_H})\rangle$ and 
$\ \langle 1({\bf \overline{16}_H})\rangle$ break $SO(10)$ to the Standard 
Model.  The electroweak VEVs arise from the combinations 
$v_u = \langle 5({\bf 10_H})\rangle$ and $v_d = \langle \overline{5}
({\bf 10_H})\rangle\cos \gamma + \langle \overline{5}({\bf 16'_H})\rangle 
\sin \gamma$, while the combination orthogonal to $v_d$ gets massive at the 
GUT scale.  As such, Yukawa coupling unification can be achieved at the GUT 
scale with $\tan \beta \sim~2~-~55$, depending upon the 
$\overline{5}({\bf 10_H}) - \overline{5}({\bf 16_H})$ mixing present for the 
$v_d$ VEV.  In addition, matter superfields appear in the following 
representations:
${\bf 16_1},\ {\bf 16_2},\ {\bf 16_3};\ {\bf 16},\ {\bf \overline{16}},\ 
{\bf 16'},\ {\bf \overline{16'}}$, ${\bf 10_1},\ {\bf 10_2}$, and ${\bf 1}$'s,
where all but the ${\bf 16_i}\ (i = 1,2,3)$ get superheavy. 

The mass matrices 
then follow from Froggatt-Nielsen diagrams \cite{fn} in which the 
superheavy fields ${\bf 16}$, ${\bf \overline{16}}$, 
${\bf 16'}$, ${\bf \overline{16'}}$, ${\bf 10_1}$, ${\bf 10_2}$, and 
${\bf 1}$ are
integrated out.  These diagrams, cf. \cite{abLMA}, correspond to the following 
effective Yukawa operators for the indicated Dirac mass matrix elements:
\begin{equation}
	\begin{array}{rl}
		&33: {\bf 16}_3 \cdot {\bf 10}_H \cdot {\bf 16_3}\\[0.05in]
		&23:  \left[{\bf 16}_2 \cdot {\bf 10}_H\right]_{\overline{16}} 
		\left[{\bf 45}_H \cdot {\bf 16}_3\right]_{16}
		/M_G\\[0.05in]
		&23: \left[{\bf 16}_2 \cdot {\bf 16}_H\right]_{10} 
		\left[{\bf 16'}_H \cdot {\bf 16}_3\right]_{10}/M_G\\[0.05in]
		&13: \left[{\bf 16}_1 \cdot {\bf 16}_3\right]_{10} 
		\left[{\bf 16}_H \cdot {\bf 16'}_H\right]_{10}
		/M_G\\[0.05in]
		&12: \left[{\bf 16}_1 \cdot {\bf 16}_2\right]_{10} 
		\left[{\bf 16}_H \cdot {\bf 16'}_H\right]_{10}
		/M_G\\[0.05in]
		&11: {\bf 16}_1 \cdot {\bf 10}_H \cdot {\bf 16}_1 \cdot
                ({\bf 1}_H)^2/M^2_G\\
	\end{array}
\label{eq:effops}	
\end{equation}
where the subscripts to the brackets indicate how the fields inside the 
brackets are contracted.
The Dirac mass matrices for the up quarks, down quarks, neutrinos and charged 
leptons are then found to be
\begin{equation}
\begin{array}{ll}
M_U = \left(\matrix{ \eta & 0 & 0 \cr
  0 & 0 & - \epsilon/3 \cr 0 & \epsilon/3 & 1\cr} \right)m_U,\
  & M_D = \left(\matrix{ 0 & \delta & \delta' e^{i\phi}\cr
  \delta & 0 & - \epsilon/3  \cr
  \delta' e^{i \phi} & \sigma + \epsilon/3 & 1\cr} \right)m_D, \\[0.3in]
M_N = \left(\matrix{ \eta & 0 & 0 \cr 0 & 0 & \epsilon \cr
        0 & - \epsilon & 1\cr} \right)m_U,\
  & M_L = \left(\matrix{ 0 & \delta & \delta' e^{i \phi} \cr
  \delta & 0 & \sigma + \epsilon \cr \delta' e^{i\phi} &
  - \epsilon & 1\cr} \right)m_D.\\
\end{array}
\label{eq:Dmatrices}
\end{equation}
The above textures give the Georgi-Jarlskog relations \cite{gj} between the 
quark and lepton GUT scale masses, 
$m^0_s \simeq m^0_\mu/3,\ m^0_d \simeq 3m^0_e$
with Yukawa coupling unification holding for $\tan \beta \sim 5$.
(Here the Dirac matrices are written with the convention that 
the left-handed fields 
label the rows and the left-handed conjugate fields label the columns.
The opposite convention was used in some earlier references for this model.
Hence the matrices here are the transpose of those given in those earlier
papers.)

All nine quark and charged lepton masses, plus the three CKM angles and CP
phase, are well-fitted with the eight input parameters (the older choice is
indicated in parentheses)
\begin{equation}
\begin{array}{rlrl}
        m_U&\simeq 113\ {\rm GeV},&\qquad m_D&\simeq 1\ {\rm GeV},\\
        \sigma&= 1.83\ (1.78),&\qquad \epsilon&=0.147\ (0.145),\\
        \delta&= 0.00946\ (0.0086),&\qquad \delta'&= 0.00827\ (0.0079),\\
        \phi&= 119.4^\circ\ (126^\circ),&\qquad \eta&= 6 \times 10^{-6}\  
		(8 \times 10^{-6}),\\
\end{array}
\label{eq:inputparam}
\end{equation}
defined at the GUT scale to fit the low scale observables after evolution 
downward from $\Lambda_{GUT}$:
\begin{equation}
\begin{array}{rlrl}
           m_t(m_t) &= 165\ {\rm GeV},\quad & m_{\tau} &= 1.777\ {\rm GeV},
                \\[0.1in]
           m_c(m_c) &= 1.23\ {\rm GeV},\quad & m_\mu &= 105.7\ 
                {\rm MeV},\\[0.1in]
           m_u(1\ {\rm GeV}) &= 3.6\ {\rm MeV}, \quad & m_e &= 0.55\ 
                {\rm MeV},\\[0.1in]
           m_b(m_b) &= 4.25\ {\rm GeV},\quad & V_{cb} &= 0.0410,\\[0.1in]
           m_s(1\ {\rm GeV}) &= 148\ {\rm MeV},\quad & V_{us} &= 0.220,\\[0.1in]
           m_d(1\ {\rm MeV}) &= 7.9\ {\rm MeV},\quad & 
                |V_{ub}/V_{cb}| &= 0.090,\\[0.1in]
           \delta_{CP} &= 64^\circ,\quad & \sin 2\beta &= 0.72.
\end{array}
\label{eq:fit}
\end{equation}
A better overall agreement with experiment \cite{pdb} is obtained with the 
new parameter values aside
from the electron mass which is most sensitive to small corrections.
With no extra phases present other than the one appearing in the CKM mixing 
matrix, the vertex of the
CKM unitary triangle occurs near the center of the presently allowed region
with $\sin 2\beta \simeq 0.72$.  The Hermitian matrices $M_U M_U^\dagger,\ 
M_D M_D^\dagger$, and $M_N M_N^\dagger$ are diagonalized with small left-handed
rotations, while $M_L M_L^\dagger$ is diagonalized by a large left-handed 
rotation.  This accounts for the small value of $V_{cb} = (U^\dagger_{U_L} 
U_{D_L})_{cb}$, while $|U_{\mu 3}| = |(U^\dagger_{L_L} 
U_{\nu_L})_{\mu 3}|$ will turn out to be 
large for any reasonable right-handed Majorana mass matrix, $M_R$ \cite{abb}.

The effective light neutrino mass matrix, $M_\nu$, is obtained from the 
type I seesaw 
mechanism once the right-handed Majorana mass matrix, $M_R$, is specified.  
The large atmospheric neutrino mixing $\nu_\mu \leftrightarrow 
\nu_\tau$ arises primarily from the structure of the charged lepton mass 
matrix $M_L$, while the structure of the right-handed Majorana mass matrix 
$M_R$ determines
the type of $\nu_e \leftrightarrow \nu_\mu,\ \nu_\tau$ solar neutrino mixing,
so that the solar and atmospheric mixings are essentially decoupled in 
the model.  The LMA solar neutrino solution is obtained with a special form of 
$M_R$, as will be seen in a moment. However, this special form can be 
explained by the structure of the Froggatt-Nielsen diagrams \cite{abLMA}.  The 
most general form for the right-handed Majorana mass matrix considered in 
\cite{abLMA} which gives the large mixing angle (LMA) solar neutrino solution is
\begin{equation}
          M_R = \left(\matrix{c^2 \eta^2 & -b\epsilon\eta & a\eta\cr
                -b\epsilon\eta & \epsilon^2 & -\epsilon\cr
                a\eta & -\epsilon & 1\cr}\right)\Lambda_R,\\
\label{eq:MR}
\end{equation}
where the parameters $\epsilon$ and $\eta$ are those introduced in 
Eq. (\ref{eq:Dmatrices}) for the Dirac sector.  With $a \neq b = c$, 
for example, the structure
of $M_R$ arises in the following way. The VEV of one particular Higgs 
singlet that has $\Delta L = 2$ 
contributes to all nine matrix elements giving a factorized
rank 1 form. The VEV of a second Higgs singlet also breaks lepton number,
but contributes only to the 13 and 31 elements of $M_R$.

Given the right-handed Majorana mass matrix above, the seesaw formula results
in 
\begin{equation}
\begin{array}{rcl}
	M_\nu &=& - M_N M^{-1}_R M^T_N\\[0.1in] 
		&=& -\left(\matrix{ 0 & 
                        \frac{1}{a-b} \epsilon & 0\cr 
                        \frac{1}{a-b} \epsilon & \frac{b^2-c^2}{(a-b)^2} 
                        \epsilon^2 
                        & \frac{b}{b-a} \epsilon\cr 0 & \frac{b}{b-a} \epsilon 
                        & 1\cr} \right)m^2_U/\Lambda_R.
\end{array}
\label{eq:MnuI}
\end{equation}
As a numerical example, with just three additional input parameters:
$a=1,\ b=c=2$ and $\Lambda_R = 2.65 \times 10^{14}$ GeV, one obtains 
\begin{equation}
 	M_\nu = -\left(\matrix{ 0 & -\epsilon & 0\cr 
                        -\epsilon & 0 & 2\epsilon\cr 0 & 2\epsilon & 1\cr}
                        \right)m^2_U/\Lambda_R,\\
\end{equation}
leading to 
\begin{equation}
\begin{array}{ll}
          \multicolumn{2}{l}{m_1 = 5.1 \times 10^{-3}\ {\rm eV},\quad m_2 = 
                9.1 \times 10^{-3}\ {\rm eV},\quad m_3 = 52 \times 10^{-3}\ 
                {\rm eV},}\\[0.05in]
          M_1 \simeq M_2 \simeq 2.3 \times 10^{8}\ {\rm GeV},\quad & M_3 = 2.7 
                \times 10^{14}\ {\rm GeV},\\[0.05in]
          \Delta m^2_{32} = 2.6 \times 10^{-3}\ {\rm eV^2},
                \quad &\sin^2 2\theta_{\rm atm} = 0.991,
                \\[0.05in]
          \Delta m^2_{21} = 5.6 \times 10^{-5}\ 
                {\rm eV^2},\quad & \tan^2 \theta_{12} = 0.48,\\[0.05in]
          U_{e3} = -0.0172-0.00094i,\quad &\sin^2 2\theta_{\rm 13} = 
		0.0012,\\[0.05in]
          J = 2.0 \times 10^{-4},\quad \delta_{CP} = 177^\circ,\quad & 
                \chi_1 = -180^\circ, \quad \chi_2 = 90^\circ,\\ 
\end{array}
\label{eq:nuoutput}
\end{equation}
to be compared with the present atmospheric, solar and reactor data  
and best-fit point in the LMA region \cite{data}
\begin{equation}
\begin{array}{rlrl}
	\Delta m^2_{32} &\simeq 2.6 \times 10^{-3}\ {\rm eV^2},&
		\quad \sin^2 2\theta_{atm}&= 1.0 \quad (\geq 0.92 {\rm\ at\ 90\% 
		\ c.l.}),\\
	\Delta m^2_{21} &= 7.1 \times 10^{-5}\ {\rm eV^2},&
                \quad \tan^2 \theta_{12} &= 0.44.\\
\label{eq:nudata}
\end{array}
\end{equation}
In fact, the whole presently-allowed LMA region can be covered with 
a thin strip in the $a - b$ plane given by $b = c \simeq 1.9 + 1.4 a$.
In this region, $\sin^2 2\theta_{13} = 0.0004 - 0.0012$.  Although the 
prediction for $U_{e3}$ is thus 3 - 5 times larger than the CKM mixing
element $V_{ub}$, a Neutrino Factory would be required to reach that range
of $\sin^2 2 \theta_{13}$.

It is interesting to remark that the hierarchy exhibited by the light 
left-handed neutrinos is a weak and normal one as is typical in $SO(10)$ 
models.  This is somewhat surprising,
for both the Dirac and Majorana neutrino matrices, $N$ and $M_R$, exhibit
strong hierarchies.  As is apparent these hierarchies nearly cancel each other
in the type I seesaw mechanism.  

It is a simple matter to compute the effective mass parameter for neutrino-less
double beta decay.  For this purpose, we first note that the MNS neutrino 
mixing matrix is given by the product of the two unitary matrices 
which diagonalize the charged lepton and light left-handed neutrino mass 
matrices, i.e., $V_{MNS} = U^\dagger_{L_L} U_{\nu_L}$.  The diagonalization 
occurs as follows:
\begin{equation}
\begin{array}{rcl}
	U^T_{L_L} M_L M^\dagger_L U^*_{L_L} &=& {\rm diag}
		(m^2_e,\ m^2_\mu,\ m^2_\tau), \\[0.1in]
	U^T_{\nu_L} M_{\nu_L} U_{\nu_L} &=& {\rm diag}(m_1,\ m_2,\ m_3), \\
\end{array}
\label{eq:diagon}
\end{equation}
where the mass eigenvalues are taken to be positive.  Clearly an arbitrary
phase transformation can be made on $U_{L_L}$ but not on $U_{\nu_L}$.  Hence
the mixing matrix $V_{MNS}$ has the form
\begin{equation}
	V_{MNS} = U_{MNS} \Phi,\quad \Phi = {\rm diag}(\exp^{i\chi_1},
		\ \exp^{i\chi_2},\ 1),
\label{eq:mns}
\end{equation}
where $U_{MNS}$ has the standard form with the Dirac CP phase appearing in
the 13 element, while $\chi_1,\ \chi_2$ are the two Majorana CP phases.  In 
terms of the above, the effective mass in neutrino-less double beta decay can
then be written as
\begin{equation}
\begin{array}{rcl}
	\langle m_{ee} \rangle &=& |\sum_j m_j \left( U_{MNS}\Phi\right)^2_{1j}|
		\\[0.1in]
		&=& |\sum_j m_j \eta_j |U_{MNS,1j}|^2|,
\end{array}
\label{eq:ee}
\end{equation}
where $\eta_j = \exp^{2i\chi_j}$ for $j = 1,2$ is the CP-parity of the 
$j$th lepton.  For the example illustrated, $\chi_1 = -180^\circ,\ 
\chi_2 = 90^\circ$ and $\langle m_{ee}\rangle = 0.57$ meV, well below the 
present or future limit of observability, typical for a normal hierarchy 
spectrum.  The lightest two neutrino states have opposite CP-parity.
	 	
Note that the two lightest heavy right-handed neutrinos are nearly degenerate
with 
\begin{equation}
	(M_2 - M_1)/M_2 = 1.21 \times 10^{-4}.\\
\label{eq:Mdegeneracy}
\end{equation}
Actually they are nearly degenerate in magnitude only, their opposite relative
signs that appear in the eigenvalues of $M_R$ signify they have exactly
opposite $CP$-parity, a feature which enables them to evolve downward from the
GUT scale without their separation receiving large radiative corrections. 
The small relative separation of the quasi-degenerate 
states $N_1$ and $N_2$ suggests that some resonance enhancement of 
leptogenesis may result from the level crossing.  As we shall see in the next 
Section, this separation is too large by more than an order of magnitude to 
produce enough leptogenesis. Moreover, at least one of the  four parameters 
$a,\ b,\ c$ and $\Lambda_R$ in the right-handed Majorana matrix defined in 
(\ref{eq:MR}) must be complex in order to generate a lepton asymmetry. 
We shall attempt to modify the parameter assignments in $M_R$ to achieve 
satisfactory leptogenesis after discussing the required conditions in the 
next Section. 

\section{Leptogenesis with Type I Seesaw}

Here we present the basic formulas for calculation of the lepton asymmetry
which results in baryogenesis with the type I seesaw mechanism.  For this 
purpose, we can use as a guide the recent phenomenological studies 
carried out by the authors of ref. \cite{eryafs}.  There is an  
important difference in our approaches, however.  They chose to work in the 
charged lepton flavor basis with the right-handed Majorana mass matrix $M_R$
diagonal, while our model was naturally developed in the $SO(10)$ flavor basis.
Hence we must transform our neutrino matrices $M_N$ and $M_R$ to the basis
in which $M_R$ is diagonal in order to apply the conventional formulas 
in refs. \cite{fy,res,eryafs}.

The basic assumption is that a lepton asymmetry $\epsilon_i$ is generated 
by the $CP$-violating out-of-equilibrium decays of the right-handed neutrino
$N_i$.  Since the right-handed Majorana mass term violates lepton number by
two units, $N_i$ is identical to its charge conjugate state and can decay
in two ways:
\begin{equation}
	N_i \rightarrow \nu_j + \phi,\ \bar{\nu}_j + \phi^\dagger;\ 
		i,j = 1,2,3\\
\label{eq:Ndecay}
\end{equation}
where $\phi$ is the Higgs field coupling the right-handed $N_i$ to the light
left-handed neutrino $\nu_j$.  The Yukawa coupling involved is given by the 
{\it ij}th element
of the  Dirac matrix $M'_N$ in the basis where the right-handed Majorana 
neutrino mass matrix $M'_R$ is diagonal.  Let us label that Yukawa coupling by 
$h'_{ij}$.  The lepton asymmetry is then traced to the imaginary part of the 
interference arising from the direct decay diagram and the one-loop diagrams 
depicted in Fig. 1:
\begin{eqnarray}
	\epsilon_i &=& \frac{\Gamma(N_i \rightarrow \nu_j\ \phi)
		- \Gamma(N_i \rightarrow \bar{\nu}_j\ \phi^\dagger)}
		{\Gamma(N_i \rightarrow \nu_j\ \phi)
		+ \Gamma(N_i \rightarrow \bar{\nu}_j\ \phi^\dagger)}\nonumber 
		\\[0.1in]
		&=& {1\over{8\pi}}\sum_{k\neq i}f\left({|M_k|^2}/{|M_i|^2}\right)
			{\cal I}_{ki},
\label{eq:epsilon}
\end{eqnarray}
where 
\begin{eqnarray}
	{\cal I}_{ki} &=& \frac{{\rm Im}\left[(h'^\dagger h')^2_{ik}\right]}
		{(h'^\dagger h')_{ii}},\\[0.1in]
	f(x) &=& \sqrt{x}\left[{1\over{1-x}} + 1 - (1+x)\log\left({{1+x}\over{x}}
		\right)\right].
\label{eq:fofx}
\end{eqnarray}	
When summed over both neutrino and charged lepton final states for all three 
families, the total lepton asymmetry is still given by the right-hand side of 
Eq. (\ref{eq:epsilon}); hence we have omitted the subscript $j$ on $\epsilon_i$.
The first term in $f(x)$ arises from the self-energy correction while the second
and third terms arise from the one-loop vertex.  
In the limit of quasi-degenerate neutrinos, $f(x)$ for $x = |M_k|^2/|M_i|^2$
reduces to 
\begin{equation}
 	\lim_{x \rightarrow 1} f(x) = -{{|M_i|}\over{2(|M_k| - |M_i|)}}.
\label{eq:fofxlimit}
\end{equation}
A resonance enhancement thus arises from the level crossing or near vanishing
of the denominator of $f(x)$ in the case of nearly degenerate masses.
The lepton asymmetry $\epsilon_1$ associated with the decay of $N_1$ thus
simplifies to 
\begin{equation}
	\epsilon_1 = - \frac{1}{16\pi}\frac{|M_1|}{|M_2| - |M_1|}{\cal I}_{21}.
\label{eq:eps1}
\end{equation}
The lepton asymmetry $\epsilon_2$ associated with the decay of $N_2$ is found
to be equal to that of $\epsilon_1$ for nearly degenerate states.  In either
case, the enhancement is limited by the decay widths of the two states for 
$||M_2| - |M_1|| \sim \Gamma_1/2 \simeq \Gamma_2/2$, cf. \cite{philaftsis}, 
where 
\begin{equation}
	\Gamma_i = {1\over{8\pi}}(h'^\dagger h')_{ii}|M_i|.
\label{eq:Gamma}
\end{equation}

In Eq. (\ref{eq:epsilon}) above, $h'$  is the Yukawa coupling matrix for 
the Dirac neutrinos in the basis where the right-handed Majorana matrix $M_R$ 
is diagonal, which we shall call the primed basis.
Denoting mass matrices written in that basis with primes, one has then
$h'= {M'_N}/(v\sin \beta)$ with $v = 174$ GeV.
In Eq. (2) the mass matrices are given in the original $SO(10)$ flavor basis, 
which we call the unprimed basis.

If the various neutrino mass matrices are diagonalized as follows
\begin{equation}
\begin{array}{rl}
   U^T_{N_L} M_N U^*_{N_R} &= {\rm diag}(m_u,\ 9m_c,\ m_t),\\[0.1in]
   U^T_{\nu_L} M_\nu U_{\nu_L} &= {\rm diag}(m_1,\ m_2,\ m_3),\\[0.1in]
   U^\dagger_{M_R} M_R U^*_{M_R} &= {\rm diag}(M_1,\ M_2,\ M_3),\\
\end{array}
\label{eq:transf}
\end{equation}
then one has that 
\begin{equation}
	M'^\dagger_N M'_N = U^T_{M_R} M^\dagger_N M_N U^*_{M_R},
\label{eq:connect}
\end{equation}
The transformation matrix $U_{M_R}$ is 
uniquely determined provided we require the mass eigenvalues in  
Eq. ({\ref{eq:transf}) be real and positive.
The factor of 9 in the diagonalized $M_N$ can be
understood by comparing $M_N$ and $M_U$ in Eq. (\ref{eq:Dmatrices}).

The lepton asymmetry produced by the decay of $N_i$ is partially diluted by 
the lepton number-violating processes themselves \cite{washout}.  This 
washout is determined as a function of an effective mass given by 
\begin{equation}
	\tilde{m}_i = (M'^\dagger_N M'_N)_{ii}/|M_i|.
\label{eq:mtilde}
\end{equation}
The washout factor $\kappa_i$ for $i = 1,2$ with $\tilde{m}_i$ in the range 
$10^{-2}\ - \ 10^3$ eV is approximated by 
\begin{equation}
	\kappa_i(\tilde{m}_i) \simeq 0.3\left({{10^{-3}{\rm eV}}\over
		{\tilde{m}_i}}\right)\left(\log {{\tilde{m}_i}\over
		{10^{-3}{\rm eV}}}\right)^{-0.6}.
\label{eq:kappa}
\end{equation}

The lepton number asymmetry produced per unit entropy at temperature 
$T > M_{1,2}$, taking into account decays of both $N_1$ and $N_2$ and their 
nearly equal washout factors, is then given by \cite{asy}
\begin{eqnarray}
	\frac{n_L}{s} &\simeq& \frac{2\kappa_1 \epsilon_1}{s} \frac{g_N T^3}{\pi^2}
		\\[0.1in]
		&=& \frac{90}{2\pi^4} \frac{g_N}{g_*} \kappa_1 \epsilon_1.
\label{eq:numasy}
\end{eqnarray}
We have used the expression for the entropy of the co-moving volume,
$s = (2/45)g_* \pi^2 T^3$.  Here $g_N = 2$ refers to the two spin degrees of 
freedom of each decaying Majorana neutrino, while $g_* = 106.75$ refers to 
the effective number of relativistic degrees of freedom contributing to the 
entropy in the absence of supersymmetric particles.  Hence we find
\begin{equation}
	(\frac{n_L}{s})^{\rm SM} \simeq 8.66 \times 10^{-3} \kappa_1 \epsilon_1.
\label{eq:SMasy}
\end{equation}

The corresponding $B - L$ asymmetry per unit entropy is just the negative of
$n_L /s$, since baryon number is conserved in the right-handed Majorana 
neutrino decays.  While $B - L$ is conserved by the electroweak interactions
following those decays, the sphaleron processes violate $B + L$ conservation 
and convert the $B - L$ asymmetry into a baryon asymmetry.  Following the 
work of Harvey and Turner \cite{ht} the connection is  
\begin{equation}
	\frac{n_B}{s} \simeq - \frac{24 + 4N_H}{66 + 13N_H}\frac{n_L}{s},
\label{eq:blconv}
\end{equation}
where $N_H$ is the number of Higgs doublets.  Again in the absence of 
supersymmetric contributions, $N_H = 1$ and the proportionality factor is 
$-28/79$.  With the entropy density $s = 7.04 n_\gamma$ in terms of the photon 
density, the baryon asymmetry of the Universe, defined by the ratio $\eta_B$  
of the net baryon number to the photon number, is given in terms of the 
lepton asymmetry $\epsilon_i$ and the washout parameter $\kappa_i$ by 
\begin{equation}
	\eta_B^{SM} \equiv \frac{n_B}{n_\gamma} \simeq - 0.0216 \kappa_1 
		\epsilon_1.
\label{eq:etaB}
\end{equation}
Successful leptogenesis will require that the 
final result for $\eta_B$ should lie in the observed range \cite{etaB}
\begin{equation}
	\eta_B = (6.15 \pm 0.25) \times 10^{-10}.
\label{eq:etaBdata}
\end{equation}

In the case where supersymmetric particles are considered to contribute to 
the decays, entropy and sphaleron interactions, the following modifications
are in order: (a) The presence of sleptons and higgsinos in the self-energy 
and vertex loops will double the interference terms without affecting 
the tree-level decay rates, so the asymmetry is doubled.
(b) The presence of sleptons and higgsinos in the decay products of 
the right-handed neutrinos will double the decay rates without affecting 
the asymmetry. (c) With the sneutrino counterparts of the decaying heavy 
neutrinos taken into account, the lepton asymmetry is further doubled.
(d) The value of the effective number of relativistic degrees of freedom
is now $g_* = 228.75$.
(e) In Eq. (\ref{eq:blconv}) we must take $N_H = 2$ for the two Higgs 
	doublets present in the supersymmetric case.

The result is that the lepton asymmetry is replaced by a factor of 4 times
the value of $\epsilon_1$ given in Eq. (\ref{eq:numasy}), but since the widths
of the decaying states are doubled, the mass separation should be taken twice
as large in the computation of the asymmetry.  The washout factor as determined
by the Boltzmann equations is also affected, but in a very complicated fashion
with the supersymmetric particles present \cite{gnrrs}.  We shall assume that 
the expression in Eq. (\ref{eq:kappa}) is still a good approximation although
this is open to question.  The appropriate expressions now read
\begin{eqnarray}
	\left(\frac{n_L}{s}\right)^{SUSY} &\simeq& 1.62 \times 10^{-2} 
        \kappa_1 \epsilon_1,
		\\[0.1in]
	\left(\frac{n_B}{s}\right)^{SUSY} &\simeq& -5.62 \times 10^{-3} 
        \kappa_1 \epsilon_1,
		\\[0.1in]
	\eta^{SUSY}_B &\simeq& -0.0396 \kappa_1 \epsilon_1.
\label{eq:susy}
\end{eqnarray}

We now apply the above framework to the $SO(10)$ model of \cite{ab,abLMA} to 
determine just how successful
leptogenesis would be in inducing baryogenesis given the mass matrices and
parameters that lead to the observed neutrino masses and mixings. 
In this model, most of the complex phases can be rotated away from the
Dirac mass matrices $M_U$, $M_D$, $M_L$, and $M_N$; in fact all but two can,
which were called $\alpha$ and $\phi$ in \cite{ab}. The phase $\alpha$
gets set to zero by fitting the quark masses which is why it does not show
up in Eq. (\ref{eq:Dmatrices}).  The phase $\phi$ is
more important and is responsible for the CP violating phase $\delta_{CKM}$
in the CKM matrix. In order to get substantial leptogenesis in this model there
must be a large phase in $M_R$. Now, in \cite{abLMA} $M_R$ was taken to be 
real, as it was in in Sect. II, simply for convenience of analysis, as the 
interest there was only atmospheric and solar neutrino oscillations rather than 
leptonic CP-violating effects.  An example of the fit with real $M_R$ is given 
in Table 1 as case (I-0).  Note that with the type I 
seesaw mechanism only four parameters, $a,\ b,\ c$ and $\Lambda_R$, are required
in addition to the eight present in the four Dirac mass matrices to obtain
excellent agreement with the neutrino mass and mixing data.  However, there 
is no reason for $M_R$ to be real, and so here we shall investigate complex 
values.  

A simple choice of complex parameters for $M_R$ is
\begin{equation}
\begin{array}{rlrl}
	a &= 1.2 - 0.45i,\qquad\qquad & b &= 2.0,\\
	c &= 2.0,	& \Lambda_R &= 2.65 \times 10^{14}\ {\rm GeV}.\\
\end{array}
\label{eq:case1}
\end{equation}
This we call case (I-1), and the resulting observables are shown also in
Table 1. Note that while the neutrino mixing parameters are all within 
the present acceptable range, the baryon asymmetry $\eta_B$ is much too 
small, as it is almost four orders of magnitude 
below the observed value.
This can be traced to the fact that the relative separation of the two
quasi-degenerate right-handed neutrino masses is too large compared
with the widths of their levels, $(M_2 - M_1)/\Gamma_1 = 34.2$.

The best results we have been able to obtain in the model arise if
we allow small non-zero values in the 12, 13, 21, and 31 entries of the 
Dirac neutrino matrix. Consider, for example, the choice
\begin{equation}
\begin{array}{ll}
	a = 0.25 + 0.15i,\qquad\qquad & b = 1.2 + 0.9i,\\
	c = 0.25 + 0.25i,  & \Lambda_R = 2.9 \times 10^{14}\ {\rm GeV},\\
	(M_N)_{12} = (M_N)_{21} = - 0.65 \times 10^{-5},\qquad & 
		(M_N)_{13} = (M_N)_{31} = - 1.0 \times 10^{-5},\\
\end{array}
\label{eq:case2}
\end{equation}
which we call case (I-2).
We have checked that the introduction of these non-zero values into the 
Dirac neutrino mass matrix, and likewise for the up quark mass matrix,
does not destroy the good agreement for the quark masses and CKM mixings.
We see from Table~1 that the neutrino mixing parameters for case (I-2) are 
also in very good agreement with the present known data.  The splitting of 
two quasi-degenerate right-handed neutrino masses has now been reduced to 
half the width of either state, which maximizes the resonance enhancement.
Note that the relative CP-parity of the two lightest neutrino states
is no longer opposite but differs from that by about $4^\circ$, a necessary
condition to get satisfactory leptogenesis.  Moreover, for this type I
seesaw model, the Dirac phase $\delta_{CP}$ is large and much closer to 
maximal than in case (I-1).  
But the baryon asymmetry is $\eta_B = 2.2 \times 10^{-10}$, which falls short 
of the observed value by a factor of three.  The biggest improvement over
case (I-1) occurs in the lepton asymmetry which has improved by three
orders of magnitude, while the washout factor is only slightly larger.

The numbers given for the three cases discussed so far for cases (I-0), (I-1), 
and (I-2) all left out the contributions from the supersymmetric particles.  
In case (I-3) of Table 1, we have taken those contributions into account. 
For this case we take  
\begin{equation}
\begin{array}{ll}
	a = 0.2 + 0.1i,\qquad\qquad & b = 1.15 + 0.9i,\\
	c = 0.25 + 0.3i,  & \Lambda_R = 2.84 \times 10^{14}\ {\rm GeV},\\
	(M_N)_{12} = (M_N)_{21} = - 0.65 \times 10^{-5},\qquad & 
		(M_N)_{13} = (M_N)_{31} = - 1.05 \times 10^{-5},\\
\end{array}
\label{eq:case3}
\end{equation}
We see from Table 1 that although $\eta_B$ given by Eq. (\ref{eq:susy}) in 
the SUSY case appears to be twice as large as in the SM case, as calculated 
from Eq. (\ref{eq:eps1}) $\epsilon_1$ is only half as large.  The net baryon
asymmetries are thus nearly equal for cases (I-2) and (I-3), provided one is
justified in using the same washout formula in both cases.  It thus appears
that, in this realistic model of quark and lepton masses and mixings,
obtaining sufficient baryon asymmetry through thermal leptogenesis is
somewhat problematic.

\section{Leptogenesis with Type III Seesaw}

In a recent paper \cite{typeIII} a new type of seesaw mechanism was proposed 
for light neutrino masses that can be implemented in grand unified theories 
based on $SO(10)$ or larger groups. This was called the type III see-saw 
mechanism and can be implemented in the model of \cite{ab,abLMA}, which we 
have been considering.
Indeed, in \cite{typeIIIab} it was argued that when implemented in 
this model, the type III see-saw can give both realistic neutrino masses and 
mixings and sufficiently large leptogenesis without fine-tuning of the forms 
of the matrices.  Here we will look at type III leptogenesis in more detail, 
giving numerical examples.

The type III seesaw involves introducing, in addition to the left- and 
right-handed neutrinos ($\nu_i$, $N_i$) contained in the ${\bf 16}_i$, three 
$SO(10)$-singlet neutrinos ${\bf 1}_i$. Thus the neutrino mass 
matrix is not $6 \times 6$ but $9 \times 9$, and is given by
\begin{equation}
W_{neut} = (\nu_i, N^c_i, S_i) \left( \begin{array}{ccc}
0 & (M_N)_{ij} & F'_{ij}u \\ (M^T_N)_{ij} & 0 & F_{ij}\Omega \\
(F'^T)_{ij}u & (F^T)_{ij}\Omega & {\cal M}_{ij} \end{array} \right) 
\left( \begin{array}{c} \nu_j \\ N^c_j \\ S_j \end{array} \right),
\label{eq:massmat9}
\end{equation}
where the $\nu_i \subset \overline{{\bf 5}}({\bf 16}_i)$ are the usual
left-handed neutrinos, the $N^c_i = {\bf 1}({\bf 16}_i)$ are conjugates of 
the usual right-handed neutrinos, and the $S_i = {\bf 1}_i$ are the 
$SO(10)$-singlet neutrinos. The index $i$ runs over families.
The $M_N$ submatrix represents the usual Dirac mass matrix contribution 
involving the doublet and singlet neutrinos in the ${\bf 16}_i$'s as in 
the previous Sect.  The new terms in the third row and third column arise 
from couplings involving the three singlet neutrinos $S_i$ as given by
the additional contributions to the Yukawa superpotential:
\begin{equation}
W_{RH\nu} = F^a_{ij} ({\bf 16}_i {\bf 1}_j) \overline{{\bf 16}}^a_H +
{\cal M}_{ij} {\bf 1}_i {\bf 1}_j.
\label{eq:superpot}
\end{equation}
The superscript $a$ distinguishes the different Higgs $\overline{\bf 16}_H$
representations, if there are more than one.  The $F'$ matrix in Eq. 
(\ref{eq:massmat9}) arises when at least one of the Higgs fields,
${\bf 5}(\overline{\bf 16}^a_H)$, gets an electroweak VEV, $u_a$, in the 
${\bf 5}$ $SU(5)$ subgroup direction.  Likewise, the $F$ matrix appears when 
at least one of the ${\bf 1}(\overline{\bf 16}^a_H)$ fields get a superheavy
VEV, $\Omega_a$, in the $SU(5)$ singlet direction.  More explicitly, we mean 
\begin{equation}
\begin{array}{rl}
	F'_{ij}u &= \sum_a F^a_{ij}u_a,\\[0.1in]
	F_{ij}\Omega &= \sum_a F^a_{ij}\Omega_a.\\
\end{array}
\end{equation}
With the submatrix ${\cal M}$ also superheavy, $W_{neut}$ can be diagonalized
to yield the $3 \times 3$ matrix in the light left-handed neutrino $\nu\nu$
block 
\begin{equation}
M_{\nu} = - M_N M_R^{-1} M_N^T - (M_N H + H^T M_N^T) \frac{u}{\Omega},
\label{eq:seesaw}
\end{equation}
where
\begin{equation}
\begin{array}{rl}
	M_R &= (F \Omega) {\cal M}^{-1} (F^T \Omega),\\[0.1in]
	H &\equiv (F' F^{-1})^T.\\
\label{eq:seesawdef}
\end{array}
\end{equation}

The first term in Eq. (\ref{eq:seesaw}) is the usual type I seesaw 
contribution.  The second term is the new type III seesaw contribution
and arises when the $\overline{\bf 16}^a_H$ fields develop an electroweak VEV, 
so $F' \neq 0$.  If the elements of the matrix ${\cal M}$ are small 
compared to those of $F\Omega$, then it is easy to see from Eqs. 
(\ref{eq:seesaw}) and (\ref{eq:seesawdef}) that the type I contribution
becomes negligible compared to the type III contribution.
In the limit that ${\cal M} = 0$ one 
sees from Eq. (\ref{eq:massmat9}) that the superheavy neutrinos have simply 
the mass term $F_{ij} \Omega (N^c_i S_j)$. That is, the $N^c_i$ and $S_i$ pair
up to form three Dirac neutrinos. On the other hand, if
${\cal M}$ is small (compared to $F \Omega$) but not zero, then 
these three Dirac neutrinos get slightly split into six eigenstates 
forming three nearly degenerate pseudo-Dirac
neutrinos. It is this fact that can be exploited to enhance leptogenesis.

In Ref. \cite{typeIIIab} it was assumed that the matrices 
$F$, $F'$ and ${\cal M}$ in the original flavor basis all have elements of 
the order
\begin{equation}
\left( \begin{array}{ccc} \lambda^2 & \lambda & \lambda \\
\lambda & 1 & 1 \\ \lambda & 1 & 1 \end{array} \right),
\end{equation}
where $\lambda \equiv \eta/\epsilon = 4.1 \times 10^{-5}$. This form is
suggested by that of $M_N$ in Eq. (\ref{eq:Dmatrices}), where the 11 
element is much smaller
than the other non-zero elements, possibly due to an Abelian flavor symmetry.
It is convenient to go to a basis where $F$ is diagonalized. This is done
by a biunitary transformation. We indicate quantities in this basis by the
tilde symbol. These are related to those in the original flavor basis by
\begin{equation}
	\tilde{F} = U^T F V,\quad N^c = U\tilde{N}^c,\quad S = V\tilde{S}.
\label{eq:transformed}
\end{equation}
Then 
\begin{equation}
	\tilde{F}\Omega = \left(\matrix{\lambda^2 F_1 & 0 & 0 \cr
		0 & F_2 & 0 \cr 0 & 0 & F_3 \cr}\right)\Omega
		= \left(\matrix{M_1 & 0 & 0 \cr 0 & M_2 & 0 \cr 
			0 & 0 & M_3 \cr}\right),
\label{eq:Ftrans}
\end{equation}
where the $F_i$ are of order unity. And
\begin{equation}
	\tilde{{\cal M}} = \left( \begin{array}{ccc}
		\lambda^2 g_{11} & \lambda g_{12} & \lambda g_{13} \\
		\lambda g_{12} & g_{22} & g_{23} \\ \lambda g_{13} & 
                g_{23} & g_{33} \end{array} \right) M_S,\qquad 
\tilde{F}' u = \left( \begin{array}{ccc}
\lambda^2 f'_{11} & \lambda f'_{12} & \lambda f'_{13} \\
\lambda f'_{21} & f'_{22} & f'_{23} \\ \lambda f'_{31} & f'_{32} & f'_{33} 
\end{array} \right) m_U.
\label{eq:F'trans}
\end{equation}
Again we assume $f'_{ij},\ g_{ij} \sim 1$. Here $M_S \ll \Omega$ in order to 
obtain three superheavy quasi-Dirac neutrino pairs.

Finally we transform the Dirac neutrino mass matrix according to 
$\tilde{M}_N = M_N U$. Because of the assumed form of $F$, the matrix $U$ has
the form
\begin{equation}
U = \left( \begin{array}{ccc} u_{11} & \lambda u_{12} & \lambda u_{13} \\
\lambda u_{21} & u_{22} & u_{23} \\ \lambda u_{31} & u_{32} & u_{33} 
\end{array} \right),
\label{eq:Utrans}
\end{equation}
where $u_{ij} \sim 1$ and by unitarity $u_{11} \simeq 1$.  The transformed
Dirac neutrino matrix then becomes 
\begin{equation}
\tilde{M}_N \cong \left( \begin{array}{ccc}
\eta u_{11} & \eta \lambda u_{12} & \eta \lambda u_{13} \\
\epsilon \lambda u_{31} & \epsilon u_{32} & \epsilon u_{33} \\
\lambda u_{31} & u_{32} & u_{33} \end{array} \right) m_U \equiv
\tilde{Y} m_U.
\label{eq:MNtrans}
\end{equation}
The type III seesaw mechanism given in Eq. (\ref{eq:seesaw}) with dominance
of the second term then yields for the light left-handed neutrino mass matrix
(recall that $\eta/\lambda = \epsilon$)
\begin{equation}
M_{\nu} \cong - \left[ \begin{array}{ccc}
2 \eta \left( \frac{u_{11} f'_{11}}{F_1} \right) & \epsilon 
\left( \frac{u_{11} f'_{21}}{F_1} \right) & \epsilon 
\left( \frac{u_{11} f'_{31}}{F_1} \right) \\ \epsilon 
\left( \frac{u_{11} f'_{21}}{F_1} \right) & 2 \epsilon \sum_j
\left( \frac{u_{3j} f'_{2j}}{F_j} \right) & 
\sum_j \left[\left( \frac{u_{3j} f'_{2j}}{F_j} \right) 
	+ \epsilon\left( \frac{u_{3j}f'_{3j}-u_{2j}f'_{2j}}{F_j} \right)\right] \\
\epsilon \left( \frac{u_{11} f'_{31}}{F_1} \right) &
\sum_j \left[\left( \frac{u_{3j} f'_{2j}}{F_j} \right) 
	+ \epsilon\left( \frac{u_{3j}f'_{3j}-u_{2j}f'_{2j}}{F_j} \right)\right] &
2 \sum_j \left( \frac{u_{3j} f'_{3j}}{F_j} 
	- \epsilon \frac{u_{2j}f'_{3j}}{F_j} \right) \end{array}
\right] \left( \frac{m_U^2}{\Omega} \right).
\label{eq:MnuIII}
\end{equation}
This clearly has a well-defined hierarchical form which is similar to the 
corresponding $M_\nu$ determined in Sect. III. for the type I seesaw mechanism.

Leptogenesis is almost exclusively produced by the decays of the lightest 
pair of the six superheavy neutrinos. Neglecting, as is justified, the mixing 
of the lightest pair of superheavy neutrinos with the two heavier pairs
of superheavy neutrinos, we find the effective
two-by-two mass matrix for the lightest pair to be 
\begin{equation} 
(\tilde{N}^c_1, \tilde{S}_1) \; \lambda^2 \left( \begin{array}{cc}
0 & F_1 \Omega  \\ F_1 \Omega  & g_{11} M_S \end{array}
\right) \left( \begin{array}{c} \tilde{N}^c_1 \\ \tilde{S}_1 \end{array} 
\right).
\end{equation}
If, as we assume, $M_S \ll \Omega$, these form an almost degenerate 
pseudo-Dirac pair, or equivalently two Majorana neutrinos with nearly equal
and opposite masses. These Majorana neutrinos are 
$N_{1\pm} \cong (\tilde{N}^c_1 \pm \tilde{S}_1)/\sqrt{2}$, with masses
$M_{1\pm} \cong \pm M_1 + \frac{1}{2} \tilde{{\cal M}}_{11} =
\lambda^2 (\pm F_1 \Omega + \frac{1}{2} g_{11} M_S)$.
These can decay into light neutrino plus Higgs boson via the term
$Y_{i \pm}(N_{1\pm} \nu_i)H$, where
\begin{equation}
Y_{i \pm} \cong (\tilde{Y}_{i1} \pm \tilde{F}'_{i1})/\sqrt{2}
\mp \frac{\tilde{{\cal M}}_{11}}{4 M_1}
(\tilde{Y}_{i1} \mp \tilde{F}'_{i1})/\sqrt{2}.
\label{eq:yukcoup}
\end{equation}
\noindent
Here $\tilde{Y}$ is the Dirac Yukawa coupling matrix given in Eq. 
(\ref{eq:MNtrans}).

It is straightforward to show that the lepton asymmetry per decay
produced by the decays of $N_{1\pm}$ is given by 
\cite{leptogen2,res}
\begin{equation}
\epsilon_1 = \frac{1}{4 \pi} \frac{{\rm Im} [\sum_j(Y_{j+} Y^*_{j-})]^2}
{\sum_j [|Y_{j+}|^2 + |Y_{j-}|^2]} f(M^2_{1+}/M^2_{1-}),
\label{eq:leptasy}
\end{equation}
\noindent
where $f(M^2_{1+}/M^2_{1-})$ comes from the absorptive part of the decay 
amplitude of $N_{\pm}$ and was given earlier in Eq. (\ref{eq:fofx}).  For the 
application here $f(M^2_{1+}/M^2_{1-}) \cong -M_1/2\tilde{{\cal M}}_{11} = 
-(F_1/2 g_1)(\Omega/M_S)$ which can be large if $M_S \ll \Omega$.
The expression for $f(M^2_{1+}/M^2_{1-})$ given above is only valid when the 
mass splitting $|M_+| - |M_-| = \tilde{{\cal M}}_{11}$ is larger than half the 
widths of the $N_{1\pm}$, which from Eq. (\ref{eq:Gamma}) are given by 
$\Gamma_{\pm} \cong \frac{1}{8 \pi} M_1
\sum_k|Y_{k \pm}|^2$. From Eqs. (\ref{eq:F'trans}), (\ref{eq:MNtrans}), and 
(\ref{eq:yukcoup}), one sees that 
$\Gamma_{1\pm} \sim \lambda^2 M_1/8 \pi$.  As we shall see with our numerical
examples, this condition that the splitting of $N_{1\pm}$ be comparable to 
or greater than their widths, is easily satisfied.

Making use of Eqs. (\ref{eq:yukcoup}) and (\ref{eq:leptasy}) one obtains
\begin{equation}
\epsilon_1 = \frac{1}{4 \pi} \frac{\sum_j(|\tilde{Y}_{j1}|^2 - 
|\tilde{F}'_{j1}|^2) {\rm Im}(\sum_k \tilde{Y}^*_{k1} \tilde{F}'_{k1})}
{\sum_j (|\tilde{Y}_{j1}|^2 + |\tilde{F}'_{j1}|^2)} f(M^2_{1+}/M^2_{1-}),
\end{equation}
\noindent
This can be evaluated in terms of the parameters of the model using Eqs. 
(\ref{eq:F'trans}) and (\ref{eq:MNtrans}), giving
\begin{equation}
\epsilon_1 \cong \frac{\lambda^2}{4 \pi} \left[ \frac{(|u_{31}|^2 -
|f'_{31}|^2) {\rm Im}(u_{31}^* f'_{31})}{|u_{31}|^2 + |f'_{31}|^2 + |f'_{21}|^2}
\right] f(M^2_{1+}/M^2_{1-}).
\end{equation}
The washout parameter is given approximately as before by Eq. (\ref{eq:kappa}) 
where now 
\begin{equation}
\tilde{m}_1 \equiv \frac{8 \pi v_u^2 \Gamma_{N_{1\pm}}}{M^2_{N1_{\pm}}}
\cong \lambda^2 \frac{v_u^2}{M_1} (|u_{31}|^2 + |f'_{31}|^2 + |f'_{21}|^2).
\end{equation}
The lepton asymmetry is then translated into the baryon asymmetry for the 
Universe by the same formulas which appeared earlier in Sect. III.

We now wish to give some numerical examples for this type III seesaw mechanism.
The first requirement, of course, is that we choose values of the parameters
that reproduce the neutrino mass and mixing data.  One could search over
the whole space of parameters, but this is a cumbersome task, as in the 
type III mechanism there are many parameters involved in the sector of
superheavy singlet neutrinos, including the parameters $u_{ij}$ and $f'_{ij}$.
A more convenient approach is to find values of $u_{ij}$ and $f'_{ij}$ that
make the matrix $M_\nu$ (given approximately in Eq. (\ref{eq:MnuIII})) have
a form close to that shown in Eq. (\ref{eq:MnuI}), since we already know
that that form can reproduce the neutrino mass and mixing data for suitable
$a,\ b$, and $c$.  First let us choose values of $u_{ij}$ that are simple
and such that $U$ in Eq. (\ref{eq:Utrans}) is unitary.  They can not be too 
simple, i.e., have too many zeros, or else there will not be sufficient 
leptogenesis.  We choose the following form for simplicity:
\begin{equation}
	U = \left( \matrix{1 & -\lambda (1 +\sqrt{2})i & \lambda \cr
		-\lambda (1+\sqrt{2})i & 1/\sqrt{2} & i/\sqrt{2} \cr 
		\lambda & i/\sqrt{2} & 1/\sqrt{2} \cr}\right),
\end{equation}
where $\lambda = \eta/\epsilon = 4.1 \times 10^{-5}$ as before.  We can make
$M_\nu$ very close numerically to the form in Eq. (\ref{eq:MnuI}) by taking
\begin{equation}
\begin{array}{rl}
	&f'_{11} \simeq 0,\ {\rm with}\ f'_{12}{\rm \ and}\ f'_{13}{\rm 
		\ arbitrary},\\[0.1in]
	&f'_{21} = - F_1/(b - a),\\[0.1in]
	&f'_{22} = - \left[(1+\sqrt{2})i + \frac{1}{2\sqrt{2}}
	\left(\frac{c^2 - a^2}{b-a}\right) -i\frac{\epsilon}{2\sqrt{2}}
	\left(\frac{c^2 - b^2}{b-a}\right) - (b-a)\frac{1-\sqrt{2}}{\sqrt{2}}
	\frac{f'_{31}}{F_1}\right]F_2/(b-a),\\[0.1in]
	&f'_{23} = - \left[1 - \frac{1}{2\sqrt{2}}i
	\left(\frac{c^2 - a^2}{b-a}\right) + \frac{\epsilon}{2\sqrt{2}}
	\left(\frac{c^2 - b^2}{b-a}\right) + (b-a)\frac{1-\sqrt{2}}{\sqrt{2}}i
	\frac{f'_{31}}{F_1}\right]F_3/(b-a),\\[0.1in]
	&f'_{31} \simeq 0,\\[0.1in]
	&f'_{32} = \frac{1}{2\sqrt{2}}\left[-i + 2(2+\sqrt{2})i\frac{f'_{31}}{F_1}
		\right]F_2,\\[0.1in]
	&f'_{33} = \frac{1}{2\sqrt{2}}\left[1 + 2\sqrt{2}\frac{f'_{31}}{F_1}
		\right]F_3,
\end{array}
\label{eq:fparam}
\end{equation}
For simplicity we take the $\tilde{\cal M}$ matrix of Eq. (\ref{eq:F'trans}) 
to be diagonal and set $g_{11} = g_{22} = g_{33} = 1$.  
Three examples are illustrated in Table 2, with the values of $f'_{21},\ 
f'_{22},\ f'_{23}$ determined from Eq. (\ref{eq:fparam}) by setting $a,\ b,\ c$ 
set equal to 1.2 - 0.45i, 2.0, 2.0, respectively, as in case (I-1) of 
Table 1; in addition $\Omega$ appearing in the type III seesaw is set equal 
to $\Lambda_R = 2.7 \times 10^{14}$ appearing in the case (I-1).  In these 
examples, $F_1,\ F_2,\ F_3$ and $M_S$ are allowed to vary.  We have considered
only the contributions to leptogenesis from the SM particles, since the 
previous results indicated little difference with the inclusion of the 
supersymmetric contributions, given the uncertainty in the corresponding
washout factor.
 
By using the type III formulas beginning with the transformed version of the
light neutrino mass matrix given in Eq. (\ref{eq:seesaw}), we clearly 
reproduce the neutrino mass and mixing results obtained with the type I seesaw 
mechanism in Table 1 for Case (I-1), where no modification of the original 
Dirac neutrino mass matrix was introduced.  In fact, these results are 
independent of the actual values taken for $F_1,\ F_2,\ F_3$ and $M_S$, and 
their values are not repeated in Table 2.  In particular, the Dirac 
CP-violating phase remains far from maximal at $172^\circ$.  On the other 
hand, one sees that successful leptogenesis and observed baryogenesis are 
easily obtained in all three examples for those four input parameters chosen.  
The masses of the lightest pseudo-Dirac pair of heavy neutrinos range from 
$4.5 \times 10^4$ GeV up to $4.5 \times 10^6$ GeV.  For this mass range, 
overproduction of gravitinos is not an issue \cite{gravitinos}.  The mass 
separation of the two Majorana neutrinos in the lightest pair is well above 
their decay widths.  Hence we see that the type III seesaw mechanism has a
much easier time simultaneously giving realistic light neutrino masses and 
mixings and successful leptogenesis.  In the realistic model we have 
examined, we see that it fails by at least a factor of three to give enough
leptogenesis with the conventional seesaw mechanism.  By contrast, with the 
type III seesaw mechanism as we have seen, the light neutrino mixing issue 
is decoupled from the issue of leptogenesis; moreover, there are ready-made 
pairs of nearly degenerate heavy neutrinos.  Consequently, there is no 
difficulty in the same realistic model getting successful leptogenesis with 
the type III seesaw mechanism.

\section{Conclusions}

In this paper we have explored the issue of resonant leptogenesis in a very 
predictive $SO(10)$ grand unified model which leads to reliable numerical 
results. Phenomenological studies \cite{eryafs} by Ellis, Raidal, and Yanagida,
as well as by Akhmedov, Frigerio, and Smirnov have previously suggested that 
successful leptogenesis can be easily obtained with resonant enhancement, 
though they had not looked at any specific realistic models in detail to be 
sure that both the leptogenesis and neutrino mass and mixing results can be
simultaneously satisfied.  In order to obtain the desired LMA solar neutrino 
solution in the model considered here, it was found that 
near degeneracy of the lightest two right-handed neutrinos was required.  
This feature neatly favors the resonant enhancement scenario for 
leptogenesis which can survive washout and be converted into the observed 
baryon excess by sphaleron interactions in thermal equilibrium above the 
critical electroweak symmetry-breaking temperature. 

We have first studied this problem in the conventional type I seesaw framework.
The seven model parameters, including one complex one, in the Dirac matrices 
are fixed by the charged lepton and quark mass and mixing data.  The lopsided
nature of the down quark and charged lepton mass matrices neatly accounts for
both the small value of the $V_{cb}$ quark mixing parameter and 
the near maximal mixing of the mu- and tau-neutrinos.  As originally
proposed, the right-handed Majorana neutrino mass matrix depended on just 
four real parameters which, in conjunction with the Dirac neutrino and charged
lepton mass matrices, then leads to a sub-maximal mixing of the solar 
neutrinos but no leptogenesis.  

By allowing three of the four $M_R$ parameters to be complex and 
introducing two additional very small parameters into the 
Dirac neutrino mass matrix, we were able to achieve a sizeable amount of 
leptogenesis; however, it falls short of the observed value by a factor of 
three.  The limiting factor is the requirement
that the heavy neutrino $M_1 - M_2$ mass separation must be comparable to 
or larger than half the decay width of either neutrino.  For this application 
the two masses are found to be of the order of $3 \times 10^8$ GeV and 
separated by  600 GeV.  In the supergravity scenario of SUSY breaking, however,
the gravitino problem which requires an upper bound on the reheating 
temperature of $T_R \aprle 10^7$ GeV at best makes this solution somewhat 
problematic. This can be alleviated, however, if SUSY breaking occurs via 
the gauge-mediated scenario.
 
We then considered a type III seesaw mechanism by which three $SO(10)$ singlet
neutrinos are added to the spectrum, so the neutrino matrix is $9 \times 9$.
With the singlet neutrinos at some large intermediate scale, the effective
``double seesaw'' results in three pairs of quasi-Dirac neutrinos in place
of three right-handed Majorana neutrinos.  In this scenario resonant-enhanced 
leptogenesis is achieved by the lightest pair of quasi-Dirac neutrinos, for 
which their mass separation is typically one hundred times the widths of the 
two states.  Hence successful leptogenesis is easily obtained.  With this 
type III seesaw the 
leptogenesis is completely decoupled from the neutrino mass and mixing issues,
so the good agreement of the latter with present-day observations is preserved
as observed earlier by us in \cite{typeIIIab}.  Here we have seen that with 
one choice of parameters the masses of the lightest neutrino pair can be 
of the order of $5 \times 10^4$ GeV, so the gravitino problem is no longer an 
issue.  The only drawback for this type III seesaw mechanism is the appearance
of a large number of new parameters needed to specify the expanded $9 \times 9$
neutrino matrix.\\

The research of SMB was supported in part by the Department of Energy under 
contract No. DE-FG02-91ER-40626.  One of us (CHA) thanks the Theory Group
at Fermilab for its kind hospitality.  Fermilab is operated by 
Universities Research Association Inc. under contract No. DE-AC02-76CH03000 
with the Department of Energy.
%
%

%
%
\newpage
\begin{table}
\caption[Table I]{Type I seesaw results for the four cases considered
	in the text.}
\vspace*{0.2in}
\begin{tabular}{ll|c|c|c|cc}
	& & Case (I-0): (SM) & Case (I-1): (SM) & Case (I-2): (SM) & Case (I-3): 
	(SUSY)\\ \hline
Input: & a & 1.0 & 1.20 - 0.45i & 0.25 + 0.15i & 0.20 + 0.10i \\[0.07in]
& b & 2.0 & 2.0 & 1.20 + 0.90i & 1.15 + 0.90i \\[0.07in]
& c & 2.0 & 2.0 & 0.25 + 0.25i & 0.25 + 0.3i \\[0.07in]
& $\Lambda_R$ (GeV) & $2.65 \times 10^{14}$\quad & $2.70 \times 10^{14}$\quad
	 & $2.90 \times 10^{14}$\quad & $2.84 \times 10^{14}$\quad \\[0.07in] 
& $(M_N)_{12},(M_N)_{21}$ & 0.0 & 0.0 & $-0.65 \times 10^{-5}$ &
	$-0.65 \times 10^{-5}$ \\[0.07in]
& $(M_N)_{13},(M_N)_{31}$ & 0.0 & 0.0 & $-1.0 \times 10^{-5}$ &
	$-1.05 \times 10^{-5}$\\[0.05in] \hline 
Output: & $M_1$ (GeV) & $2.31 \times 10^8$ & $2.16 \times 10^8$ & 
	$3.06 \times 10^8$ & $3.08 \times 10^8$ \\[0.07in]
& $M_2$ (GeV) & $2.31 \times 10^8$ & $2.16 \times 10^8$ & $3.06 \times 10^8$
	& $3.08 \times 10^8$ \\[0.07in]
& $M_3$ (GeV) & $2.71 \times 10^{14}$ & $2.76 \times 10^{14}$ 
	& $2.90 \times 10^{14}$ & $2.86 \times 10^{14}$ \\[0.07in]
& $\Delta M_{21}/M_2$ & $1.21 \times 10^{-4}$ & $1.31 \times 10^{-4}$ 
	& $1.96 \times 10^{-6}$ & $3.87 \times 10^{-6}$ \\[0.07in]
& $\Gamma_1/M_1$ & $3.83 \times 10^{-6}$ & $3.83 \times 10^{-6}$
	& $3.84 \times 10^{-6}$ & $3.83 \times 10^{-6}$ \\[0.07in]
& $m_1$ (meV) & 5.1 & 5.3 & 2.8 & 2.8 \\[0.07in]
& $m_2$ (meV) & 9.1 & 9.8 & 8.8 & 8.8 \\[0.07in]
& $m_3$ (meV) & 52.0 & 52.0 & 51.0 & 52.0 \\[0.07in]
& $\Delta m^2_{32}\ {\rm (eV^2)}$ & $2.6 \times 10^{-3}$ & $2.6 \times 10^{-3}$
	& $2.5 \times 10^{-3}$  & $2.6 \times 10^{-3}$ \\[0.07in]
& $\Delta m^2_{21}\ {\rm (eV^2)}$ & $5.6 \times 10^{-5}$ & $6.9 \times 10^{-5}$
	& $6.9 \times 10^{-5}$ & $6.9 \times 10^{-5}$ \\[0.07in]
& $\sin^2 2\theta_{atm}$ & 0.991 & 0.988 & 0.979 & 0.981 \\[0.07in]
& $\sin^2 2\theta_{sol}$ & 0.87 & 0.86 & 0.80 & 0.80 \\[0.07in]
& $\tan^2 \theta_{12}$ & 0.48 & 0.46 & 0.39 & 0.39 \\[0.07in]
& $\sin^2 2\theta_{13}$ & 0.0012 & 0.0006 & 0.0012 & 0.0014 \\[0.07in]
& $|U_{e3}|$ & 0.017 & 0.012 & 0.018 & 0.019 \\[0.07in]
& $\delta_{CP}$ & $177^\circ$ & $172^\circ$ & $76^\circ$ & $71^\circ$ 
	\\[0.07in]
& $\chi_1,\ \chi_2$ & $-180^\circ,\ 90^\circ$ & $24^\circ,\ 117^\circ$ 
	& $102^\circ,\ -170^\circ$ & $102^\circ,\ -170^\circ$ \\[0.07in]
& $\langle m_{ee}\rangle$ (meV) & 0.57 & 0.58 & 0.44 & 0.44 \\[0.07in]
& $\epsilon_1$ & 0.0 & $-6.4 \times 10^{-7}$ & $-1.2 \times 10^{-3}$  
	& $-6.1 \times 10^{-4}$ \\[0.07in]
& $\tilde{m}_1$ (eV) & 12.6 & 13.5 & 9.5 & 9.5 \\[0.07in]
& $\kappa_1$ & $6.2 \times 10^{-6}$ & $5.8 \times 10^{-6}$ & 
	$8.3 \times 10^{-6}$ & $8.4 \times 10^{-6}$ \\[0.07in]
& $\eta_B$ & 0.0 & $0.80 \times 10^{-13}$ & $2.2 \times 10^{-10}$  
	& $2.0 \times 10^{-10}$ \\
\end{tabular}
\end{table}
\newpage
\begin{table}
\caption[Table 2]{Type III seesaw results for three examples patterned 
	after case (I-1) in Table~1.  The light neutrino mass and mixings
	results obtained are identical to those in case (I-1) of Table~1
	and are not repeated here.}
\vspace*{0.2in}
\begin{tabular}{ll|c|c|c}
	& & Case (III-1): (SM)\qquad & Case (III-2): (SM)\qquad & Case (III-3): 
		(SM) \qquad\\[0.1in] \hline\hline
Input: & $\Omega$ (GeV) & $2.7 \times 10^{14}$ & $2.7 \times 10^{14}$ & $2.7 
	\times 10^{14}$\\[0.07in] 
& $F_1$ & 1.0 & 10. & 0.1 \\[0.07in]
& $F_2$ & 1.0 & 0.1 & 0.1 \\[0.07in]
& $F_3$ & 1.0 & 1.0 & 1.0 \\[0.07in]
& $M_S$ & $4.3 \times 10^5$ & $8.5 \times 10^8$ & $1.0 \times 10^5$ \\[0.07in]
& $f_{11}/F_1$ & 0.0		 & 0.0		 & 0.0		\\[0.07in]
& $f_{12}/F_2$ & 1.0		 & 1.0		 & 1.0		\\[0.07in]
& $f_{13}/F_3$ & 1.0		 & 1.0		 & 1.0		\\[0.07in]
& $f_{21}/F_1$ & -0.950 + 0.534i & -9.496 + 5.341i & -0.095 + 0.053i\\[0.07in]
& $f_{22}/F_2$ & -2.279 - 1.537i & -0.227 - 0.154i & -0.228 - 0,154i\\[0.07in]
& $f_{23}/F_3$ & -0.194 + 1.523i & -0.194 + 1.523i & -0.194 + 1.523i\\[0.07in]
& $f_{31}/F_1$ & 0.0		 & 0.0		 & 0.0		\\[0.07in]	
& $f_{32}/F_2$ & -0.354i	 & -0.035i		 & -0.035i	      \\[0.07in]
& $f_{33}/F_3$ & 0.354		 & 0.354		 & 0.354		\\[0.07in]
& $g_{11}$	 & 1.0		 & 1.0		 & 1.0		\\[0.07in]
& $g_{22}$	 & 1.0		 & 1.0		 & 1.0		\\[0.07in]
& $g_{33}$	 & 1.0		 & 1.0		 & 1.0		\\[0.05in] 
	\hline 
Output: & $M_1$ (GeV) & $\pm 4.50 \times 10^5 $ & $\pm 4.50 \times 10^6$ 
	& $\pm 4.50 \times 10^4$\\[0.07in]
& $M_2$ (GeV) & $\pm 2.70 \times 10^{14}$ & $\pm 2.70 \times 10^{13}$
	& $\pm 2.70 \times 10^{13}$\\[0.07in]
& $M_3$ (GeV) & $\pm 2.70 \times 10^{14}$ & $\pm 2.70 \times 10^{14}$ 
	& $\pm 2.70 \times 10^{14}$\\[0.07in]
& $(M_{1+} + M_{1-})/M_{1+}$ & $1.6 \times 10^{-9}$ & $3.15 \times 10^{-7}$ 
	& $3.7 \times 10^{-9}$\\[0.07in]
& $\Gamma_1/M_1$ & $6.9 \times 10^{-11}$ & $3.88 \times 10^{-9}$ 
	& $3.82\times 10^{-11}$\\[0.07in]
& $\chi_1,\ \chi_2$ & $-156^\circ,\ 113^\circ$ & $24^\circ,\ -67^\circ$ & 
	$-156^\circ,\ 113^\circ$ \\[0.07in]
& $\epsilon_1$ & $2.5 \times 10^{-5}$ 
	& $1.6 \times 10^{-4}$ & $-1.4 \times 10^{-4}$\\[0.07in]
& $\tilde{m}_1$ (eV) & 0.10 & 0.57 & 0.48\\[0.07in]
& $\kappa_1$ & $1.2 \times 10^{-3}$ & $1.8 \times 10^{-4}$
	& $2.1 \times 10^{-4}$\\[0.07in]
& $\eta_B$ & $-6.2 \times 10^{-10}$ 
	& $-6.1 \times 10^{-10}$ & $6.3 \times 10^{-10}$\\
\end{tabular}
\end{table}
%
%
\begin{figure}
\vspace*{0.5in}
\center{
\epsfbox{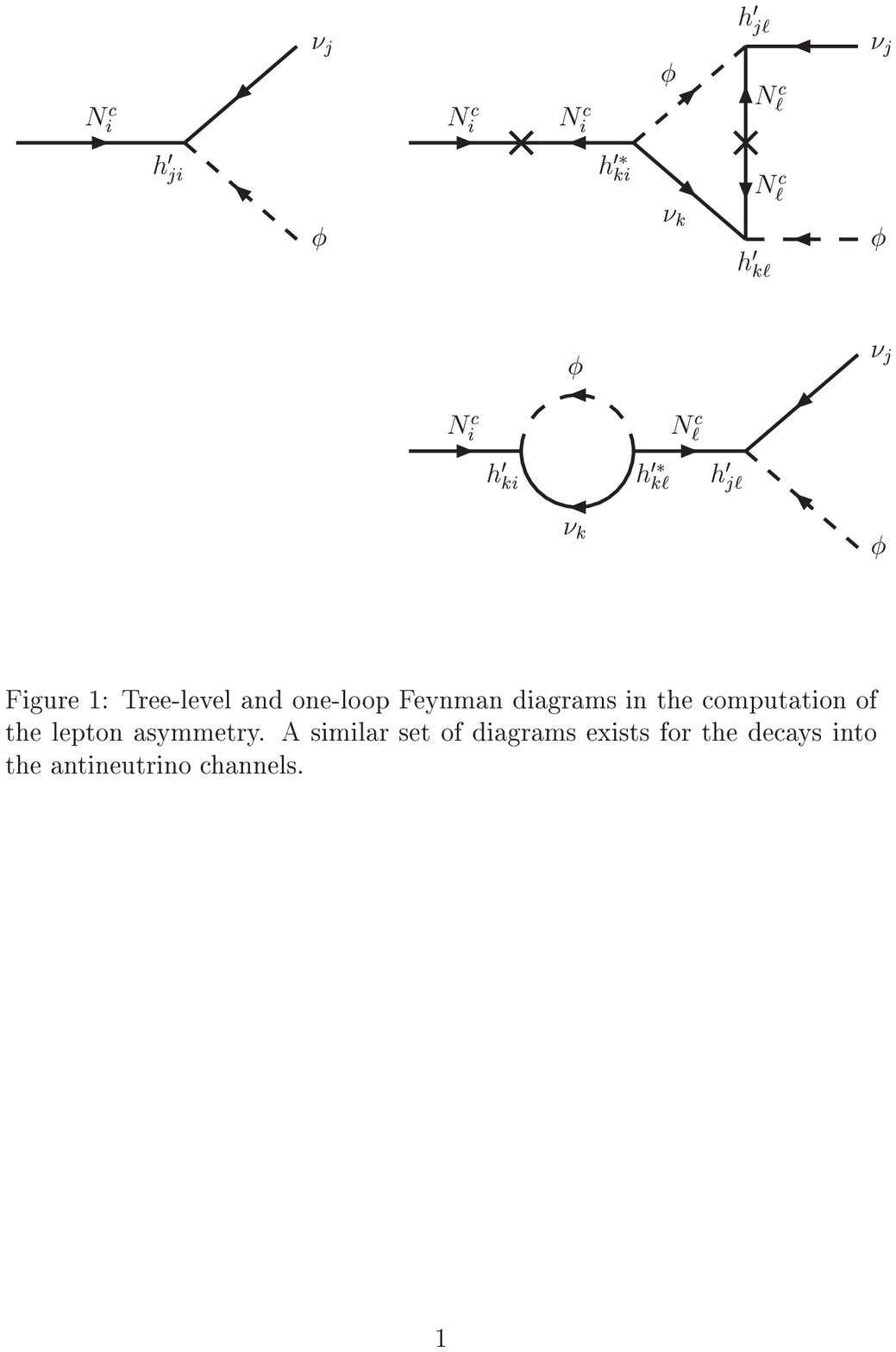}
}
\vspace {0in}
\end{figure}

\newpage
\begin{thebibliography}{99}
\vspace*{-0.1in}
%
\bibitem{models} A sampling of some still successful grand unified models 
	with their early references include the following:
	C.H. Albright and S.M. Barr, Phys. Rev. D {\bf 58}, 013002 (1998);
	K.S. Babu, J.C. Pati, and F. Wilczek, Nucl. Phys. B {\bf 566}, 33 (2000);
	Mu-C. Chen and K.T. Mahanthappa, Phys. Rev. D {\bf 62}, 113007 (2000);
	R. Kitano and Y. Mimura, Phys. Rev. D {\bf 63}, 016008 (2001);
	M. Bando and N. Maekawa, Prog. Theor. Phys. {\bf 106}, 1255 (2001);
	T. Fukuyama and N. Okada, JHEP {\bf 0211}, 011 (2002);
	M. Bando and M. Obara, Prog. Theor. Phys. {\bf 109}, 995 (2003);
	S.F. King and G.G. Ross, Phys. Lett. B {\bf 574}, 239 (2003);
	H.S. Goh, R.N. Mohapatra, and S.-P. Ng, Phys. Rev. D {\bf 68}, 115008
		(2003);
	M. Bando, S. Kaneko, M. Obara, and M. Tanimoto, Phys. Lett. B {\bf 580},
		229 (2004);
	C.S. Aulakh, B. Bajc, A. Melfo, G. Senjanovic, and F. Vissani, 
		{\it ibid.} {\bf 588}, 196 (2004). 

\bibitem{sakharov}  A.D. Sakharov, Pisma Zh. Eksp. Teor. Fiz. {\bf 5},
			32 (1967) [JETP Lett. {\bf 5}, 24 (1967)].	

\bibitem{ht}  J.A. Harvey and M.S. Turner, Phys. Rev. D {\bf 42}, 3344 (1990).
			
\bibitem{fy}  M. Fukugita and T. Yanagida, Phys. Lett. B {\bf 174}, 45 (1986).

\bibitem{res} M. Flanz, E.A. Paschos, U. Sarkar, and J. Weiss, Phys. Lett.
		B {\bf 389}, 693 (1996); L. Covi and E. Roulet, {\it ibid.}
		{\bf 399}, 113 (1997).

\bibitem{eryafs} J. Ellis, M. Raidal, and T. Yanagida, Phys. Lett. B {\bf 546},
		228 (2002); E.Kh. Akhmedov, M. Frigerio, and A.Yu. Smirnov, 
		JHEP {\bf 0309}, 021 (2003).

\bibitem{pu} A. Pilaftsis and T.E.J. Underwood, Nucl. Phys. {\bf B692}, 303
		(2004).

\bibitem{abb} C.H. Albright, K.S. Babu, and S.M. Barr, Phys. Rev. 
        Lett. {\bf 81}, 1167 (1998).

\bibitem{ab}  C.H. Albright and S.M. Barr, Phys. Rev. Lett. {\bf 85}, 244
		(2000); Phys. Rev. D {\bf 62}, 093008 (2000).

\bibitem{msw}  L. Wolfenstein, Phys. Rev. D {\bf 17}, 2369 (1978); S.P. 
	Mikheyev and A.Yu. Smirnov, Yad. Fiz {\bf 42}, 1441 (1985) [Sov. J. 
	Nucl. Phys. {\bf 42}, 913 (1985)].

\bibitem{mns} B. Pontecorvo, Sov. Phys.JETP {\bf 6}, 429 (1957); {\it ibid.},
	{\bf 7}, 172 (1958); Z. Maki, M. Nakagawa, and S. Sakata, Prog. Theor.
	Phys. {\bf 28}, 870 (1962).

\bibitem{typeIII} S.M. Barr, Phys. Rev. Lett. {\bf 92}, 101601-1 (2004).

\bibitem{typeIIIab} C.H. Albright and S.M. Barr, Phys. Rev. D {\bf 69},
	073010 (2001).

\bibitem{br} S.M. Barr and S. Raby, {\it Phys. Rev. Lett.} {\bf 79}, 
        4748 (1997).

\bibitem{fn}  C.D. Froggatt and H.B. Nielsen, Nucl. Phys. B {\bf 147},
	277 (1979).

\bibitem{abLMA} C.H. Albright and S.M. Barr, Phys. Rev. D {\bf 64}, 073010
	(2001).

\bibitem{gj} H. Georgi and C. Jarlskog, {\it Phys. Lett.} {\bf B86}, 297
        (1979).

\bibitem{pdb} Particle Data Group, K. Hagiwara {\it et al.}, Phys. Rev. 
	D {\bf 66}, 010001 (2002).

\bibitem{data} G.L. Fogli, E. Lisi, A. Marrone, and D. Montanino, Phys. Rev.
	D {\bf 67}, 093006 (2003); M.B. Smy {\it et al.}, Super-Kamiokande Coll.,
	{\it ibid.} {\bf 69}, 011104(R) (2004).
	
\bibitem{philaftsis} A. Pilaftsis, Phys. Rev. D {\bf 56}, 5431 (1997).

\bibitem{washout} E.W. Kolb and M.S. Turner, {\it The Early Universe} 
	(Addison-Wesley, Redwood City, CA, 1990); A. Pilaftsis, Int. J. Mod.
	Phys. A {\bf 14}, 1811 (1999); M. Flanz and E.A. Paschos, Phys. Rev.
	D {\bf 58}, 113009 (1998).

\bibitem{asy} See, eg., L. Covi, E. Roulet, and F. Vissani, Phys. Lett. 
	B {\bf 384}, 169 (1996).

\bibitem{etaB} D.N. Spergel {\it et al.}, Astrophys.J. Suppl. {\bf 148},
	175 (2003).

\bibitem{gnrrs}  G.F. Giudice, A. Notari, M. Raidal, A. Riotto, and 
	A. Strumia, Nucl. Phys. {\bf B685}, 89 (2004).

\bibitem{leptogen2} M.A. Luty, Phys. Rev. D {\bf 45}, 455 (1992); 
	W. Buchm\"{u}ller and T. Yanagida, Phys. Lett. B {\bf 302}, 240 (1993);
	H. Murayama and T. Yanagida, {\it ibid.}, {\bf 322}, 349 (1994);
	R. Jeannerot, Phys. Rev. Lett. {\bf 77}, 3292 (1996).

\bibitem{gravitinos} M. Kawasaki, K. Kohri, and T. Moroi, astro-ph/0402490.
\end{thebibliography}
\end{document}